\documentclass[aps, pra, twocolumn, reprint]{revtex4-1}

\usepackage{amsmath}    
\usepackage{mathrsfs}
\usepackage{graphicx}   
\usepackage{verbatim}   
\usepackage{subfigure}  
\usepackage{upgreek}
\usepackage{hyperref}
\begin{document}

\newcommand{\TEMOO}{TEM$_{00}~$}
\newcommand{\TEMIO}{TEM$_{10}~$}
\newcommand{\SiN}{Si$_3$N$_4~$}\
\newcommand{\um}{$\upmu$m}
\newcommand{\opt}{_{\text{opt}}}
\newcommand{\mat}{_{\text{mat}}}
\newcommand{\h}{\mathscr{H}}
\newcommand{\CCC}{\mathscr{C}}
\newcommand{\SSS}{\mathscr{S}}

\title{Enhanced optomechanical levitation of minimally supported dielectrics}
\author{Tina M\"uller}
\author{Christoph Reinhardt}
\author{Jack C. Sankey}

\affiliation{McGill University Department of Physics}

\date{\today}

\begin{abstract}
Optically levitated mechanical sensors promise isolation from thermal noise far beyond what is possible using flexible materials alone. One way to access this potential is to apply a strong optical trap to a minimally supported mechanical element, thereby increasing its quality factor $Q_m$. Current schemes, however, require prohibitively high laser power ($\sim10$ W), and the $Q_m$ enhancement is ultimately limited to a factor of $\sim 50$ by hybridization between the trapped mode and the dissipative modes of the supporting structure. Here we propose a levitation scheme taking full advantage of an optical resonator to reduce the \emph{circulating} power requirements by many orders of magnitude. Applying this scheme to the case of a dielectric disk in a Fabry-Perot cavity, we find a tilt-based tuning mechanism for optimizing both center-of-mass and torsional mode traps. Notably, the two modes are trapped with comparable efficiency, and we estimate that a $10$-\um-diameter, $100$-nm-thick Si disc could be trapped to a frequency of $\sim 10$ MHz with only $30$ mW circulating in a cavity of (modest) finesse $1500$. Finally, we simulate the effect such a strong trap would have on a realistic doubly-tethered disc. Of central importance, we find torsional motion is comparatively immune to $Q_m$-limiting hybridization, allowing a $Q_m$ enhancement factor of $\sim1500$. This opens the possibility of realizing a laser-tuned $10$ MHz mechanical system with a quality factor of order a billion.
\end{abstract}

\maketitle

A central theme in optomechanics is to use the forces exerted by light to enable new functionality in mechanical systems of all sizes \cite{Kippenberg2007Cavity, Aspelmeyer2014Cavity, 2014Cavity}. For example, laser radiation has been used to cool the motion of flexible solids to the quantum ground state \cite{Teufel2011Sideband, Chan2011Laser} at which point quantum motion becomes apparent in the optical spectrum \cite{Safavi2012Observation, arxivPurdy2014Optomechanical, arxivLee2014Observation}. Equally impressively, ultrathin membrane ``microphones'' have been made sensitive enough to detect the ``hiss'' from the quantized nature of incident laser light \cite{Purdy2013Observation}, and the mechanical response to this noise has been shown to squeeze the light \cite{Brooks2012Non,Safavi2013Squeezed,Purdy2013Strong}. Furthermore, it now seems a realistic goal to create optomechanical force detectors capable of ``sensing'' delicate superpositions of forces from a variety of quantum systems, and faithfully imprinting this information on an arbitrary wavelength of light \cite{Tian2010Optical,Stannigel2010Optomechanical,Regal2011From, Safavi2011Proposal, Wang2012Using}, as supported by demonstrations of wavelength conversion in the classical regime \cite{Hill2012Coherent, Liu2013Electromagnetically, Andrews2014Bidirectional}. 

All of these (and other sensing) pursuits are at some level fundamentally limited by the dissipation of the mechanical element; any channel by which energy escapes is also a channel by which the thermal environment applies unwanted force noise, limiting measurement sensitivity and destroying coherence. One approach to circumvent this limitation is to use an optically-levitated dielectric particle as the mechanical element. Because laser light can be made to exert a minuscule radiation pressure force noise, such systems are predicted to achieve an unprecedented level of coherence \cite{Chang2010Cavity, Romero2010Toward, Singh2010All}, enabling ultrasensitive force / mass detection and quantum optomechanics experiments in a room temperature apparatus. Promising work toward purely levitated systems is currently underway \cite{Li2010Measurement, Neukirch2013Observation, Kiesel2013Cavity, Gieseler2014Nonlinear}. 

A complementary approach is to only \emph{mostly} replace the material supports with light, by applying a strong optical trap to a mechanical element fabricated with minimal material supports. By storing a large fraction of its mechanical energy in the light field, the quality factor $Q_m$ can in principle be increased beyond the limits imposed by the material \cite{Chang2012Ultrahigh}. The idea can be understood by imagining a simple harmonic oscillator of mass $m$ and material spring constant $K\mat$ stiffened by an essentially dissipationless optical spring $K\opt$. Assuming material dissipation enters as the imaginary component of $K\mat$ \cite{Saulson1990Thermal}, the equation of motion is
\begin{equation}\label{eq:basic}
m \partial^2_t x + (K\mat +K\opt) x = 0.
\end{equation}
This results in an optically tuned mechanical frequency $\omega_m$$=$$\sqrt{\omega\mat^2+\omega\opt^2}$ (with $\omega\mat = \sqrt{K\mat/m}$ and $\omega\opt = \sqrt{K\opt/m}$) and a quality factor $Q_m$ enhanced by a factor $K\opt/K\mat = \omega_m^2/\omega\mat^2$. Importantly, $Q_m \propto \omega_m^2$, meaning not only does $Q_m$ increase with frequency, but the overall dissipation rate also \emph{decreases}, and the mass experiences a reduced thermal force noise from the material near $\omega_m$. If this noise can be made insignificant compared with that of the trapping light, such a system would essentially behave as though it is optically levitated. Note that here we ignore the quantum noise contribution of the laser light, focusing instead on the elimination of coupling to the thermal bath. The effects of radiation pressure shot noise (RPSN) have already been well-analyzed in the context of levitation \cite{Chang2010Cavity,Romero2010Toward,Chang2012Ultrahigh}, and could be either nulled out with a single-port optical resonator in the ``resolved sideband'' regime \cite{Thompson2008Strong} (i.e.\ for low-noise applications), or enhanced with a two-port resonator \cite{Miao2009Standard} or the ``bad cavity'' limit (for RPSN measurements and squeezing applications). We briefly revisit these ideas in Section \ref{sec:summary}.

\begin{figure}[htb]
	\includegraphics[width=0.40\textwidth]{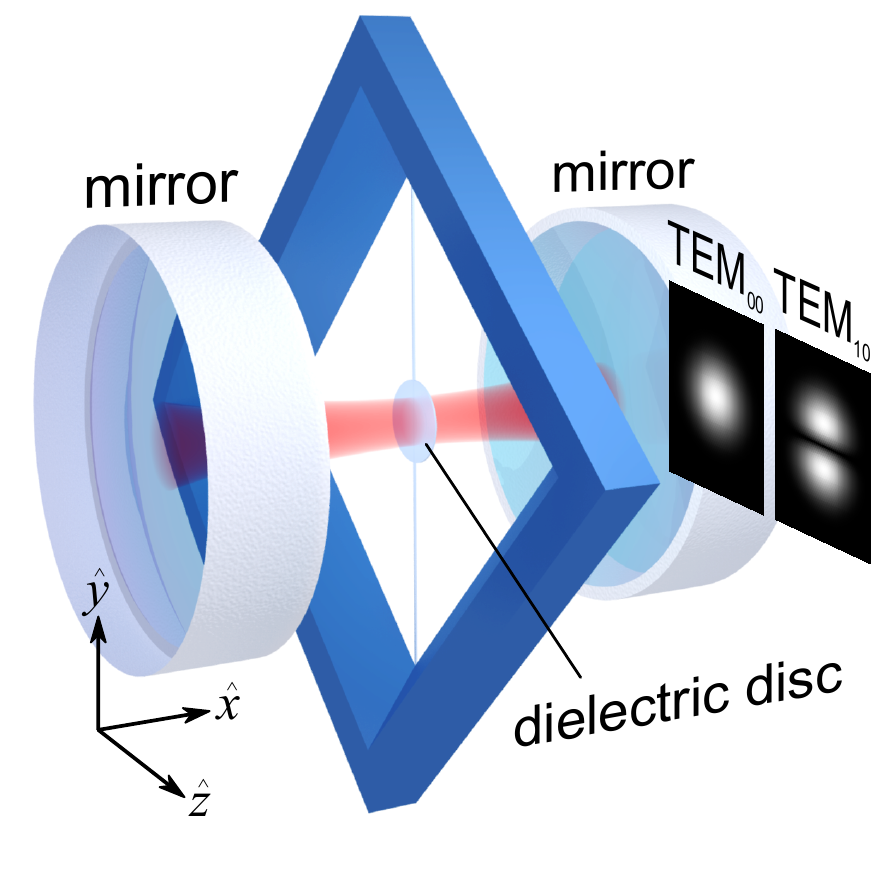}
	\caption{A proposed levitation geometry, comprising a doubly-tethered dielectric disc immersed in a cavity optical field. Inset shows the intensity profile of the cavity's \TEMOO and \TEMIO modes relative to the structure.}
	\label{fig:intro}
\end{figure}

The primary advantage of this ``partially levitated'' approach is that the mechanical element can be fabricated in a variety of shapes using standard lithography, and attached to a manageable frame (e.g.\ as in Fig.\ \ref{fig:intro}). This firstly eliminates the need for launch-and-trap techniques, and secondly enables a finer level of control over the device's orientation with respect to the light field -- this is of central importance for increasing the efficiency of the optical trap, as discussed below. Finally, a wide array of optically-incompatible probes (e.g.\ sharp / scattering tips, nanomagnets, etc) could be fabricated on regions of the device lying outside the optical field, for coupling to external systems such as qubits \cite{Stannigel2010Optomechanical} or nuclear spins \cite{Degen2009Nanoscale}. 

Initial work with a singly-tethered SiO$_2$ disc ($10$ \um~$\times$~130 nm) trapped by a high-power retro-reflected standing wave has provided encouraging confirmation of the physics described above \cite{Ni2012Enhancement}. The quality factor increased $\propto\omega_m^2$ until a peak enhancement factor $\sim 50$, at which point it plummeted due to the first of two practical limitations: in this geometry, the ``violin string'' modes of the tether hybridize with the trapped center-of-mass motion whenever their frequencies become nearly degenerate, introducing a second dissipation channel. The violin modes could be staved off by shortening or stiffening the tether, but this would simultaneously increase $K\mat$, leading to the second limitation: the laser power required to achieve $K\opt \gg K\mat$ would become unreasonably large ($>10$~W) and the material would eventually melt. 

The following article addresses both issues, first providing an efficient cavity levitation scheme based on quadratic optomechanical coupling \cite{Bhattacharya2008Optomechanical, Thompson2008Strong, Jayich2008Dispersive, Sankey2009Improved, Sankey2010Strong, Rosenberg2010Sensitive, Hill2011Mechanical, Flowers2012Fiber, Lee2015Multimode}, and second suggesting a torsional geometry (Fig.\ \ref{fig:intro}) that is far less susceptible to $Q_m$-limiting mechanical hybridization. Section \ref{sec:general-quadratic} reviews the general features of quadratic coupling within the context of levitation, and discusses how the quality factor (or finesse) of an optical resonator can be exploited beyond the simple amplification of input light. Sections \ref{sec:general-levitation} - \ref{sec:specific-levitation} then apply the scheme to the case of a dielectric disc in a Fabry-Perot cavity, and we derive expressions for both center-of-mass and torsional trap efficiencies. Of note, we find the efficiencies are comparable, and that they can be increased by a factor of order the cavity finesse by properly orienting the disc. Finally, in Section \ref{sec:comsol} we simulate the response of a readily-fabricated device to this newly accessible trap strength. In particular, we find the $Q_m$ associated with torsional motion can be increased by more than three orders of magnitude. 

\section{Quadratic Coupling and Trapping}\label{sec:general-quadratic}
When two nearly-degenerate optical modes having linear (dispersive) optomechanical coupling also scatter into one another, the resulting hybridized mode can exhibit a purely quadratic optomechanical coupling. Here we review this type of coupling within the context of optical levitation. As discussed below, whether on chip \cite{Rosenberg2010Sensitive, Hill2011Mechanical}, in a fiber cavity \cite{Flowers2012Fiber}, or in a macroscopic cavity \cite{Thompson2008Strong, Jayich2008Dispersive, Sankey2009Improved, Sankey2010Strong, Flowers2012Fiber, Karuza2013Tunable, Lee2015Multimode}, the ability to fully utilize the quality factor $Q_\gamma$ (or finesse $F$) of an optical resonator relies on the ability to control the scattering rate between the underlying optical modes. 

\begin{figure}[htb]
	\includegraphics[width=0.40\textwidth]{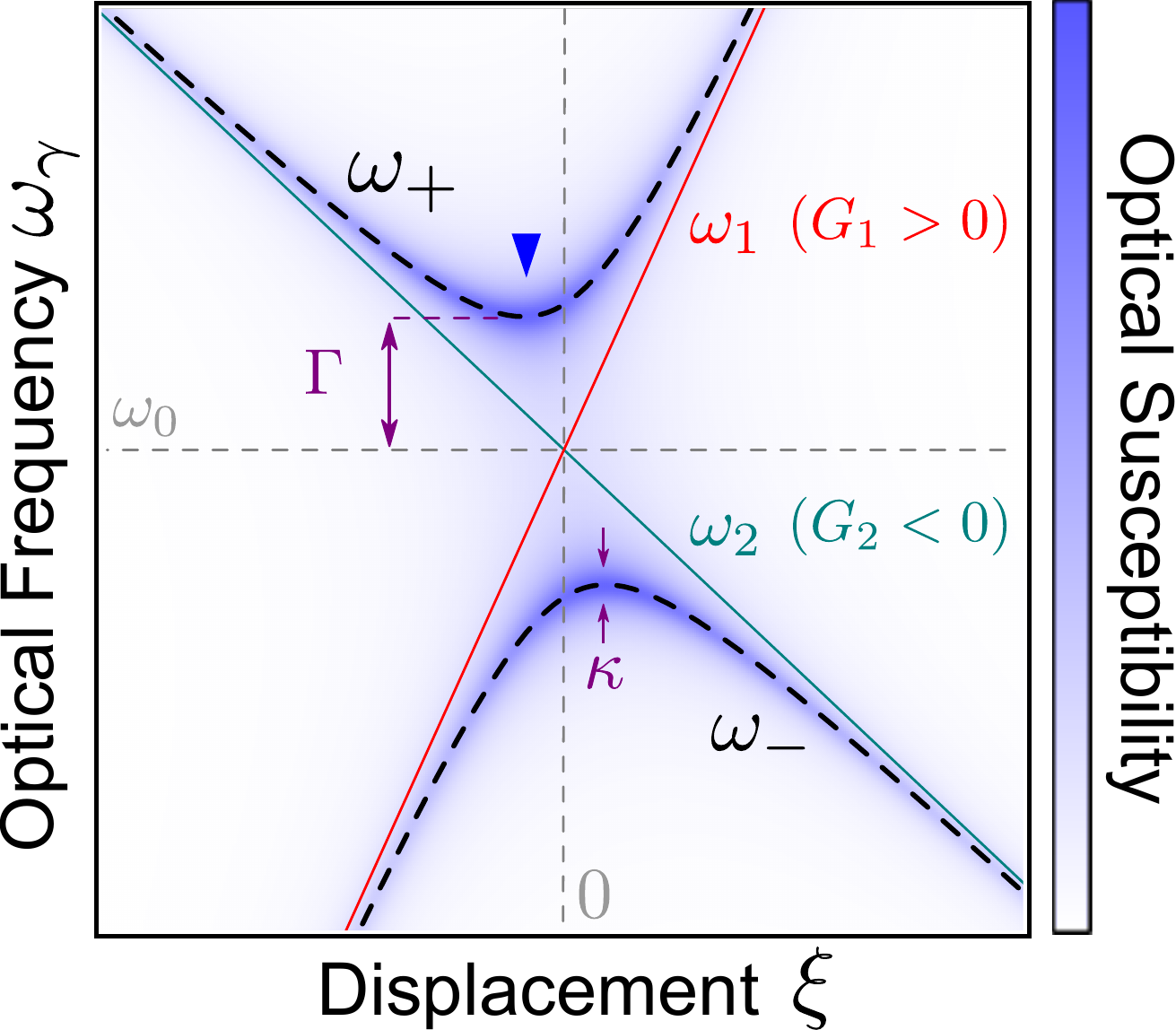}
	\caption{Quadratic optomechanical coupling. When two optical modes of linear coupling constants $G_1$ (red) and $G_2$ (green) are degenerate (at $\xi=0$ and $\omega_\gamma = \omega_0$), an inter-mode scattering rate $\Gamma$ leads to adiabatic frequencies $\omega_\pm$ (black dashed curves). The underlying blue gradient represents the optical (amplitude) susceptibility of the mixed modes, with linewidth $\kappa$ set by the decay rate of the cavity. Blue arrow indicates an optical mode with positive quadratic coupling $\partial_\xi^2\omega_+>0$, which can be used to generate a stable optical trap.}
	\label{fig:crossing}
\end{figure}

The idea of using quadratic coupling to trap a mechanical element's center of mass has been discussed for some time. The per-photon mechanical frequency shift appears in the Hamiltonians of Refs.\ \cite{Thompson2008Strong, Bhattacharya2008Optomechanical}, was identified as a stable trapping mechanism in Ref.\ \cite{Bhattacharya2008Optomechanical}, and generalized to the case of non-adiabatic cavity response in Ref.\ \cite{Jayich2008Dispersive}. 

To illustrate the concept for any type of motion, consider the generic cavity spectrum drawn in Fig.\ \ref{fig:crossing}: a mechanical element's displacement coordinate $\xi$ linearly changes the frequency of two optical modes such that their frequencies are $\omega_1 = \omega_0 + G_1 \xi$ (red line) and $\omega_2 = \omega_0 + G_2 \xi$ (green line), with constants $G_1$, $G_2$, and degenerate frequency $\omega_0$ at $\xi=0$. If the optical modes also scatter into one another at a rate $\Gamma$ (note this can be via any mechanism, not necessarily the mechanical element itself), the resulting eigenmodes will have $\xi$-dependent frequencies (dashed curves) given by \cite{Jayich2008Dispersive}
\begin{equation}\label{eq:wpm-general}
\begin{aligned}
\omega_{\pm}&=\omega_{0}+G_{+}\xi\pm\sqrt{G_{-}^{2}\xi^{2}+\Gamma^{2}} \\
&\approx\omega_{0}+G_{+}\xi \pm \left[ \Gamma +\left(G_{-}^{2}/2\Gamma\right) \xi^{2}\right]
\end{aligned}
\end{equation}
near $\xi=0$, with $G_{\pm} \equiv (G_1 \pm G_2)/2$, and an avoided gap $2\Gamma$. Assuming the optical mode responds adiabatically at the mechanical frequency $\omega_m$ (i.e.\ $\omega_m \ll \Gamma$), each photon populating the upper (+) branch will have an energy $U_+(x) = \hbar\omega_+(\xi)$, thereby exerting a static force $-\hbar G_+$ and an optical spring constant $K_+ \approx \hbar G_-^2/\Gamma$. To maximize this per-photon restoring force, one therefore engineers (or tunes) $G_1$ and $G_2$ to be of opposite sign and as large as possible, and for the scattering rate $\Gamma$ to be as small as possible.

Of course, $\Gamma$ cannot be made arbitrarily small. First, the adiabatic assumption breaks down when $\Gamma \sim \omega_m$, leading to (i) an appreciable lag in the restoring force and (ii) a larger fraction of the cavity light responding linearly, rather than quadratically. Bounding $\Gamma \gtrsim \omega_m$ correspondingly bounds the per-photon trap efficiency to
\begin{equation}
K_{+} \lesssim \hbar G_-^2 / \omega_m
\end{equation} 
where $\omega_m$ is the \emph{trapped} mechanical frequency. Near and beyond this limit, interesting new effects arise, such as enhanced quadratic readout \cite{Ludwig2012Enhanced} and Landau-Zener-St\"{u}ckelberg dynamics \cite{Heinrich2010Photon}.

A lag in the restoring force represents a particularly important concern, because it leads to instability via anti-damping. To give a sense of scale, for a spring delay $t_d$, the equation of motion is $\partial_t^2 \xi(t) = -\omega_m^2 \xi(t-t_d)$, leading to a parasitic anti-damping rate $\approx \omega_m^2 t_d/2$ in the high-$Q_m$ limit. For practical systems this can be quite significant, especially when compared with the intrinsic damping of a typical mechanical element: even if $t_d$ is of order of the round trip time for light in a 3-cm cavity, the anti-damping rate exceeds 1 kHz for a levitated 1 MHz oscillator; this is many orders of magnitude greater than the linewidth associated with a high-$Q_m$ oscillator. However, for quadratic coupling, as predicted in Ref.\ \cite{Jayich2008Dispersive} and observed in Ref.\ \cite{Lee2015Multimode} (Figs.\ 2 and 3), this anti-damping can be mitigated while still achieving a high $K_{+}$ by slightly detuning the laser and/or moving away from the purely-quadratic point. In other words, a small amount of laser cooling can prevent this lag from increasing the effective temperature. Hence, (similar to schemes incorporating dissipative optomechanics \cite{Tarabrin2013Anomalous, Sawadsky2015Observation} a significant advantage of quadratic levitation is that it does not require a second stabilizing laser, as is the case for linear optomechanical traps \cite{Corbitt2007An}. Further, in the limit $\Gamma/\omega_m \rightarrow \infty$, this trapping scheme should not suffer from static bistability \cite{Bhattacharya2008Optomechanical}.

A second lower bound on $\Gamma$ is the cavity decay rate $\kappa$ (labeled in Fig.\ \ref{fig:crossing}). If the gap $2\Gamma \lesssim \kappa$, the upper and lower branches are no longer distinct and the system reduces to a mechanical element linearly coupled to two independent optical modes. Bounding $\Gamma \gtrsim \kappa/2$ places a second upper bound on the per-photon trap efficiency
\begin{equation}\label{eq:K-Q-limit}
K_{+} \lesssim 2 \hbar G_{-}^{2}Q_\gamma/\omega_{0} = 2\hbar G_{-}^{2}F/\omega_{FSR}
\end{equation}
where $\omega_{FSR} = \pi c/ L$ is the free spectral range for the case of a Fabry-Perot resonator of length $L$. This illustrates how an optimized scattering rate can utilize the quality factor or finesse of an optical resonator to improve trap efficiency. We emphasize the upper bound is proportional to the power stored inside the cavity, so the enhancement is in addition to the ``usual'' resonant amplification of the input field. This contrasts the result obtained for thin dielectrics immersed in a standing wave, be it generated by two lasers, retro-reflected, or contained within a cavity. As pointed out in Ref.\ \cite{Ni2012Enhancement}, all three schemes generate the same per-watt trap as a free-space standing wave, even if the cavity in the latter case has a very high finesse.

\section{Optimal Levitation of a Thin Dielectric Disc}\label{sec:general-levitation}
We now describe how a thin dielectric disc positioned within a Fabry-Perot optical cavity (Fig.\ \ref{fig:intro}) can generate an optimal trap for both its center-of-mass (CM) and torsional mode (TM) motion. 

The Hermite-Gaussian modes $\phi_{\eta\mu\nu}(x,y,z)$ provide a natural orthonormal basis for the electric field in an optical cavity with curved end mirrors \cite{Siegman1986Lasers}. These modes are indexed by one longitudinal ($\eta$) and two transverse ($\mu$ and $\nu$) integers counting the number of nodes along the $\hat{x}$, $\hat{y}$, and $\hat{z}$ axes respectively. Following first-order optical perturbation theory \cite{2014Cavity}, the disc's refractive index $n$ modifies the free space Helmholtz equation as $\nabla^2 \psi + (\omega^2 / c^2)(1+V)\psi = 0$, with a perturbation ``potential'' $V(x, y, z) = n^2-1$ that is non-zero only inside the disc. If $V$ perturbs two of the basis modes $\phi_1$ and $\phi_2$ (where subscript $i$ corresponds to indices $\eta_i\mu_i\nu_i$ for brevity) so that they become nearly degenerate, the resulting eigenmodes $\psi \approx a_1\phi_1+a_2\phi_2$ and eigenfrequencies $\omega$ satisfy
\begin{equation}
\left(\begin{array}{cc}
V_{11}+1-\omega_1^2/\omega^2 & V_{12}\\
V_{21} & V_{22}+1-\omega_2^2/\omega^2
\end{array}\right)\left(\begin{array}{c}
a_1\\
a_2
\end{array}\right)=0,
\label{eq:matrix}
\end{equation}
where 
\begin{equation}\label{eq:Vij}
V_{ij} = V_{ji} \equiv \iiint V\phi_i\phi_j dx dy dz,
\end{equation} 
and $\omega_{i}$ are the unperturbed cavity mode frequencies. The diagonal matrix elements $V_{11}$ and $V_{22}$ describe the frequency shift in the absence of hybridization, while the off-diagonal elements $V_{12} = V_{21}$ mediate hybridization when the modes are nearly degenerate. This system is readily solved, and in the small perturbation limit $|\omega-\omega_{i}| \ll \omega_{i}$ the eigenvalues are
\begin{equation}
\omega_{\pm}\approx\frac{\omega_{1}^{\prime}+\omega_{2}^{\prime}}{2}\pm\sqrt{\left(\frac{\omega_{1}^{\prime}-\omega_{2}^{\prime}}{2}\right)^{2}+\left(\frac{\omega_{1}V_{12}}{2}\right)^{2}},
\label{eq:wpm}
\end{equation}
where $\omega_j^\prime \approx \omega_j(1-V_{jj}/2)$ are the perturbed eigenfrequencies in the absence of hybridization. Comparing with Eq.\ \ref{eq:wpm-general}, the scattering rate $\Gamma = \omega_1 |V_{12}|/2$. 

Typically, the integrals $V_{ij}$ must be solved numerically, but an analytical solution containing all of the relevant physics can be found with a few simplifying approximations. First, assuming the disc is positioned near cavity mode waist, wherein the wavefronts are approximately flat,
\begin{equation}\label{eq:phi}
\phi_{\eta\mu\nu} \approx \frac{H_{\mu}(Y)H_{\nu}(Z)e^{-(Y^{2}+Z^{2})/2}}{\sigma\sqrt{2^{\mu+\nu-2}\mu!\nu!\pi L}}\cos\left(kx+\pi \eta/2\right).
\end{equation}
Here, $H_j$ is the $j^\text{th}$ Hermite polynomial, $Y = \sqrt{2}y/\sigma$ and $Z = \sqrt{2}z/\sigma$ are the transverse coordinates normalized by the cavity mode radius $\sigma$, $L$ is the cavity length, and $k \approx 2\pi/\lambda$ is the effective longitudinal wave number at wavelength $\lambda$. Nominally $\sigma$ and $k$ depend on $(\eta,\mu,\nu)$ due to diffraction, but near the waists of well collimated, nearly degenerate modes, these corrections have little effect. Second, if the disc tilt angles $\theta_y$ and $\theta_z$ about the $\hat{y}$ and $\hat{z}$ axes (respectively) are small, the bounds of the $\hat{x}$-integral in $V_{ij}$ are approximately $x_0 + \theta_z y+\theta_y z \pm t_\theta/2$, where $x_0$ is the position of the disc center, and $t_\theta = t/\cos(\theta_y)\cos(\theta_z)$ is the tilt-corrected thickness along $\hat{x}$ (for actual thickness $t$). Finally, we assume the disc radius $r > \sigma$, and approximate the transverse integrals by taking their limits to infinity. Relaxing this last approximation adds additional prefactors (involving error functions) that reduce the perturbation, but this does not alter the symmetry of the problem or the tilt-based tuning mechanism discussed below. 

In this limit, an expansion of $V_{ij}$ to second order in the small quantity $t/L$ yields
\begin{equation}\label{eq:V12}
\hspace*{-0.2cm}\begin{aligned}
&V_{ij} \approx \alpha\left\{ \cos\frac{\pi(\eta_{i}-\eta_{j})}{2}\right.\\&+\tau\cos\left[2kx_{0}+\frac{\pi(\eta_{i}+\eta_{j})}{2}\right]\mathscr{C}_{\mu_{i}\mu_{j}}\left(\Theta_{z}\right)\mathscr{C}_{\nu_{i}\nu_{j}}\left(\Theta_{y}\right)\\&-\tau\cos\left[2kx_{0}+\frac{\pi(\eta_{i}+\eta_{j})}{2}\right]\mathscr{S}_{\mu_{i}\mu_{j}}\left(\Theta_{z}\right)\mathscr{S}_{\nu_{i}\nu_{j}}\left(\Theta_{y}\right)\\&-\tau\sin\left[2kx_{0}+\frac{\pi(\eta_{i}+\eta_{j})}{2}\right]\mathscr{S}_{\mu_{i}\mu_{j}}\left(\Theta_{z}\right)\mathscr{C}_{\nu_{i}\nu_{j}}\left(\Theta_{y}\right)\\&-\left.\tau\sin\left[2kx_{0}+\frac{\pi(\eta_{i}+\eta_{j})}{2}\right]\mathscr{C}_{\mu_{i}\mu_{j}}\left(\Theta_{z}\right)\mathscr{S}_{\nu_{i}\nu_{j}}\left(\Theta_{y}\right)\right\} 
\end{aligned}
\end{equation}
where we have defined normalized angles $\Theta_{y,z} = \sqrt{2}k\sigma\theta_{y,z}$, perturbation strength $\alpha = (n^{2}-1)t_{\theta}/L$, thickness correction $\tau = 1-k^{2}t_{\theta}^{2}/6$, and transverse overlap integrals
\begin{equation}
\begin{aligned}
\mathscr{C}_{nm}(\Theta)&=\int_{-\infty}^{\infty}\h_n(\chi)\h_m(\chi)e^{-\chi^{2}}\cos(\Theta\chi)d\chi\\
\mathscr{S}_{nm}(\Theta)&=\int_{-\infty}^{\infty}\h_n(\chi)\h_m(\chi)e^{-\chi^{2}}\sin(\Theta\chi)d\chi
\end{aligned}\label{eq:curlyCS}
\end{equation}
with ``normalized'' Hermite polynomials
\begin{equation}\label{eq:normalized}
\h_n(\chi) = \frac{H_n(\chi)}{\sqrt{2^n n! \pi^{1/2}}}
\end{equation}
(i.e.\ defined so that $\int \h_n\h_me^{-\chi^2}d\chi=\delta_{nm}$). Note that if $n-m$ is even, $\h_n(\chi)\h_m(\chi)$ is an even function, and $\mathscr{S}_{nm}$ vanishes by symmetry. Similarly, if $n-m$ is odd, $\mathscr{C}_{nm}$ vanishes. As a result, at least three of the terms in Eq.\ \ref{eq:V12} vanish for any pair of modes. Further, for a given set of indices $n$ and $m$, equation \ref{eq:curlyCS} can in practice be solved exactly by noticing that the complex sum
\begin{equation}
\hspace*{-0.2cm}\begin{aligned}
\mathscr{C}_{nm}+i\mathscr{S}_{nm} &= \sum_j A_j\int \chi^{j}e^{-\chi^{2}}e^{i\Theta\chi}d\chi \\
&= \sum_j A_j e^{-\Theta^{2}/4} \int (\chi^\prime+i\Theta/2)^{j}e^{-\chi^{\prime 2}}d\chi^\prime\label{eq:integral}
\end{aligned}
\end{equation}
for constant coefficients $A_j$ determined by the Hermite polynomials. Since each real term in the expansion of $(\chi^\prime+i\Theta/2)^m$ contains only \emph{even} powers of $\Theta$ and each imaginary term contains only \emph{odd}, $\mathscr{C}$ and $\mathscr{S}$ must have the form
\begin{equation}
\begin{aligned}\label{eq:exact}
\mathscr{C}_{nm}(\Theta)&= e^{-\Theta^{2}/4}C_{nm}(\Theta)\\
\mathscr{S}_{nm}(\Theta)&= e^{-\Theta^{2}/4}S_{nm}(\Theta)
\end{aligned}
\end{equation}
where $C_{nm}$ ($S_{nm}$) is an even (odd) polynomial.

Prior to specifying a particular pair of modes (Section \ref{sec:specific-levitation}), we can make basic arguments about general behavior of any two modes by further inspecting Eq.\ \ref{eq:V12}. For example, all matrix elements $V_{ij}$ exhibit a sinusoidal dependence on $x_0$ with period $\lambda/2$, and for diagonal elements $V_{jj}$ all but the first two terms in Eq.\ \ref{eq:V12} vanish. Hence, for an aligned disc ($\theta_y=\theta_z=0$) $\CCC_{jj}=1$, and $\omega_j^\prime$ oscillates as a function of $x_0$ with amplitude $\alpha\tau\omega_j$ regardless of $j$. By Eq.\ \ref{eq:exact}, tilting the disc reduces these oscillations in a mode-dependent fashion, allowing one to split nominally degenerate transverse modes via $\theta_y$ or $\theta_z$.

We can also deduce the scattering rate $\Gamma \propto |V_{ij}|$ for a small tilt by inspecting the symmetry of the integrals $\mathscr{C}$ and $\mathscr{S}$. For example, if we select a pair of modes that are one free spectral range apart ($\eta_j=\eta_i-1$) with the same transverse profile along $\hat{z}$ ($\nu_1=\nu_2$, e.g.\ the \TEMOO and \TEMIO modes in Fig.\ \ref{fig:intro}), the first, third, and fifth terms of Eq.\ \ref{eq:V12} immediately vanish. Then (assuming $\mu_j>\mu_i$ without loss of generality),
\begin{equation}\label{eq:leading}
\hspace*{-0.3cm}\begin{aligned}
\mathscr{C}_{\mu_i\mu_j} &\approx \sqrt{\frac{\mu_j!}{2^{\mu_j-\mu_i} \mu_i!}}\frac{(-\Theta)^{\mu_j-\mu_i}}{(\mu_j-\mu_i)!} \text{ for even } \mu_j-\mu_i \\
\mathscr{S}_{\mu_i\mu_j} &\approx \sqrt{\frac{\mu_j!}{2^{\mu_j-\mu_i} \mu_i!}}\frac{(-\Theta)^{\mu_j-\mu_i}}{(\mu_j-\mu_i)!} \text{ for odd }  \mu_j-\mu_i  
\end{aligned}
\end{equation} 
to leading order (see Appendix I), and both are identically zero otherwise. Importantly, $V_{ij} \propto \Theta^{\mu_i-\mu_j}$ for any $\mu_i$ and $\mu_j$, meaning the scattering rate between modes of differing $\mu$ is \emph{zero} for $\theta_z=0$ (regardless of $\theta_y$), and guaranteed to be tunable via $\theta_z$. For the simplest case of the \TEMOO and \TEMIO modes discussed below, the scattering rate $\Gamma$ is simply proportional to $\theta_z$, motivating the incorporation of two tethers as drawn in Fig.\ \ref{fig:intro}; in the absence of either (or both!), this critical orientation would be quite difficult to define.
\begin{figure*}[htb]
	\includegraphics[width=0.8\textwidth]{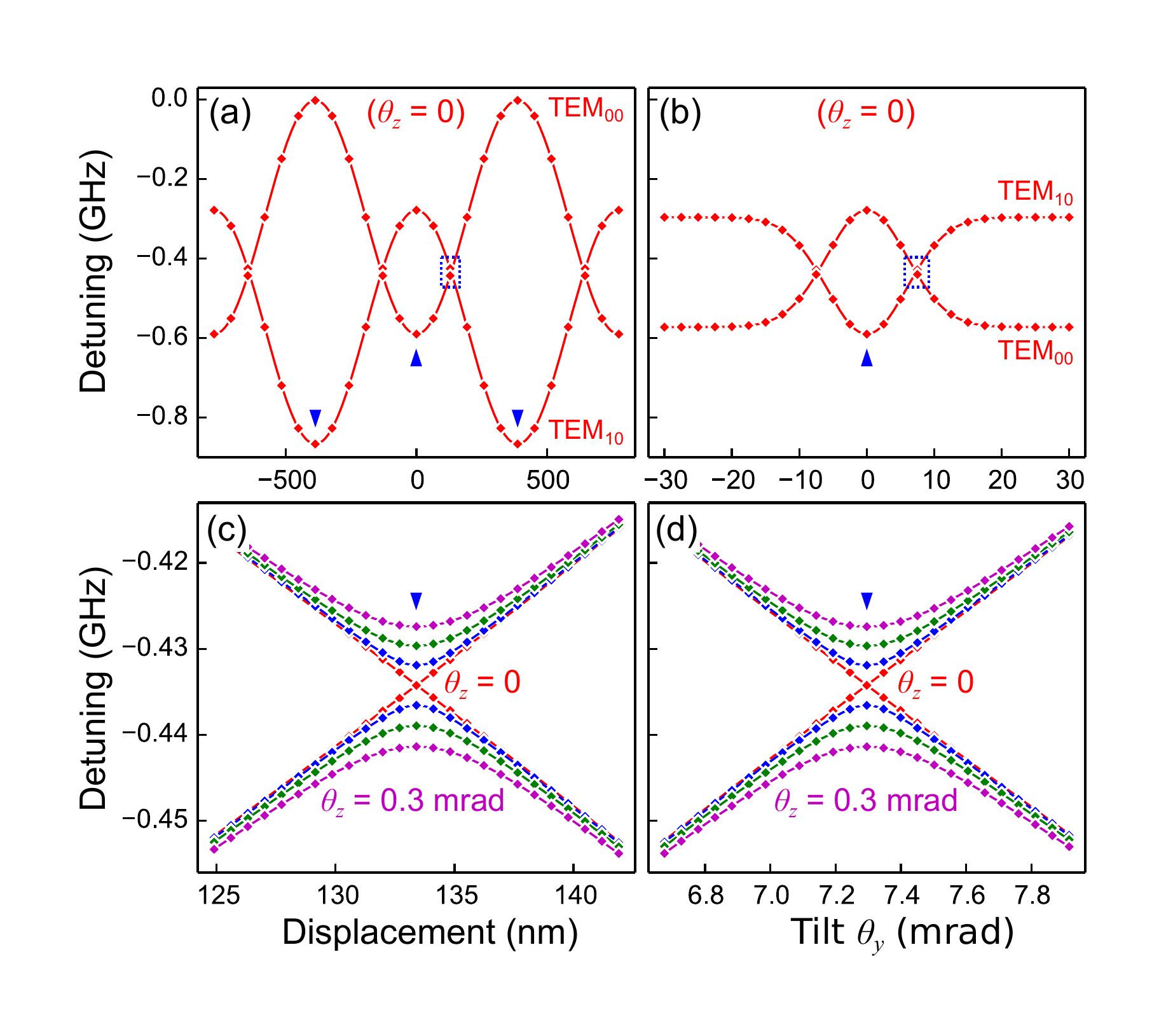}
	\caption{Single mode and enhanced cavity traps. The cavity comprises two mirrors with radius of curvature 2.5 cm separated by a length $L=4.9$ cm, and a 50-nm-thick \SiN membrane ($n=2$) positioned near the waist ($x_0=0$). Blue arrows indicate stable cavity traps, solid curves show the analytical theory, and diamonds show a numerical solution with no approximations. (a) Detuning of \TEMOO and \TEMIO cavity modes versus displacement $x_0$ for an aligned ($\theta_z=\theta_y=0$) membrane. (b) Detuning of the same modes versus tilt about the $\hat{y}$ (tether) axis for $x_0=0$. (c) and (d) show refined plots of the regions indicated by dotted boxes in (a) and (b) for fixed tilts $\theta_z =$ 0, 0.1, 0.2, 0.3 mrad about $\hat{z}$. In both cases the gap is tuned linearly with $\theta_z$.}
	\label{fig:cavity}
\end{figure*}
\section{Example: a \TEMOO-\TEMIO Crossing}\label{sec:specific-levitation}
To see how the aforementioned control can be achieved in practice, we now consider the simplest two transverse modes that can cross in a favorable way, namely  (see Fig.\ \ref{fig:intro}) the \TEMOO ($\mu_1=\nu_1=0$) and \TEMIO ($\mu_2=1$, $\nu_2=0$) modes, separated by one free spectral range ($\eta_2=\eta_1-1$). In this case, by Eqs.\ \ref{eq:integral} and \ref{eq:exact}, $C_{\mu_1\mu_1} = 1$, $C_{\mu_2\mu_2} = 1-k^2\sigma^2\theta_z^2$ ($\equiv \beta$), by Eq.\ \ref{eq:leading} $\SSS_{\mu_1\mu_2}(\Theta)\approx -k\sigma\theta_z$, and Eq.\ \ref{eq:V12} reduces to
\begin{equation}\label{eq:Vspecific}
\begin{aligned}
V_{11}&= \alpha\left\{ 1+     \tau e^{-k^{2}\sigma^{2}(\theta_{y}^{2}+\theta_{z}^{2})/2}\cos2kx_{0}\right\}\\ 
V_{22}&= \alpha\left\{ 1-\beta\tau e^{-k^{2}\sigma^{2}(\theta_{y}^{2}+\theta_{z}^{2})/2}\cos2kx_{0}\right\}\\
V_{12}&= \alpha\tau k\sigma\theta_{z}e^{-k^{2}\sigma^{2}(\theta_{y}^{2}+\theta_{z}^{2})/2}\cos2kx_{0}
\end{aligned}
\end{equation}
where $\eta_1$ is chosen (rather arbitrarily) to be a multiple of 4. Figures \ref{fig:cavity} (a) and (b) show the resulting eigenfrequencies (Eq.\ \ref{eq:wpm}) versus $x$ and $\theta_y$ for a 50-nm-thick \SiN disc ($n=2.0$) aligned in a cavity of length $L=4.9$ cm and mirror radius of curvature $2.5$ cm. Solid lines show the analytical solution, and diamonds show a ``sanity check'' numerical solution with none of the approximations beyond Eq.\ \ref{eq:wpm}. Note these expressions are valid over a much wider range of tilts than those of Ref.\ \cite{Sankey2009Improved}, sufficient to not only describe the torsional trap of interest, but also how a purely-\emph{quartic} coupling \cite{Sankey2010Strong} can be generated with just two cavity modes (see Appendix II).

Blue arrows indicate a positive quadratic dependence on $x$ or $\theta_y$, which can be used to generate stable optical traps for CM or TM motion. At these points, the ``single-mode'' (TEM$_{00}$) spring constants are
\begin{eqnarray}
K_{\text{CM},1}&=&4LP\alpha\tau k^{2}/c \label{eq:cm-spring}\\
K_{\text{TM},1}&=&LP\alpha\tau k^{2}\sigma^{2}/c \label{eq:tor-spring},
\end{eqnarray}
where $P$ is the circulating power in the cavity. We reiterate that for a thin, weakly-perturbing disc, Eq.\ \ref{eq:cm-spring} is identical to the expression derived from a 1D scattering / transfer matrix approach \cite{Ni2012Enhancement}, though a 1D theory can of course not describe torsional motion. Also notice that $K_{\text{CM},1}$ and $K_{\text{TM},1}$ differ only by a factor $\sigma^2/4$, which, together with the disc mass $m$ and moment of inertia $I \approx \frac{1}{4}mr^{2}$, results in a ratio of TM and CM optical trap frequencies 
\begin{equation}
\frac{\omega_{\text{TM}}}{\omega_{\text{CM}}} = \frac{\sigma}{r}.
\end{equation}
Importantly, this factor is of order unity when $\sigma \sim r$ (though it never exceeds unity due to the breakdown of the approximate integral bounds), so torsional motion can be trapped with an efficiency comparable to that of the center of mass. The reduction in TM trap efficiency for smaller $\sigma$ can be understood as arising from the non-uniform spring constant density across the disc: mechanical modes having more displacement near the center of the disc (where the intensity is higher) experience a stronger integrated restoring force. In the opposite ``large spot'' limit $\sigma\gg r$ this factor approaches unity at the expense of a reduced trapping efficiency for both CM and TM. The effect of finite spot size on the resulting $Q_m$-enhancement is addressed in Section \ref{sec:comsol}.

Figures \ref{fig:cavity} (c) and (d) show the crossings in more detail, for $\theta_z$ between 0 to 0.3 milliradians. In both cases the avoided gap can be tuned linearly with $\theta_z$, as expected.

In principle, the crossing points can also be tuned (e.g.\ via cavity length) to occur where the slope of the uncoupled optical modes is nearly maximal. In Fig.\ \ref{fig:cavity}, for example, $L$ was chosen because the CM and TM crossings simultaneously occur near their maximal slopes, and varying $L$ from this value allows one to shift \TEMIO vertically with respect to TEM$_{00}$. Assuming the crossings have been tuned to the optimal points, the maximum enhancement of the spring constants $K_{\text{CM},2}$ and $K_{\text{TM},2}$ for this 2-mode scheme is
\begin{eqnarray}
\frac{K_{\text{CM},2}}{K_{\text{CM},1}}&=& \frac{\omega_{FSR}}{\Gamma}\frac{(n^{2}-1)\tau t_\theta}{\lambda}\\
\frac{K_{\text{TM},2}}{K_{\text{TM},1}}&=&\frac{\omega_{FSR}}{\Gamma}\frac{(n^{2}-1)\tau t_\theta}{\lambda e}
\end{eqnarray}
where $\omega_{FSR} = \pi c/L$. For the present example, a gap $\Gamma/2\pi \sim 1$ MHz corresponds to a factor of order $\sim 200$ increase in the per-photon restoring force compared to the single-mode scheme shown in Fig.\ \ref{fig:cavity} (a-b). Since this enhancement scales as the ratio $\omega_{FSR}/\Gamma$, it is apparently (perhaps not surprisingly) beneficial to use a shorter cavity at fixed $\Gamma$. A 1 mm cavity, for example, could achieve an enhancement factor of order $\sim 15,000$.

Other simple materials exerting a larger perturbation, e.g.\ a 110-nm-thick Si disc (see Appendix III) or a high-reflectivity structured dielectric \cite{Bui2012High, Kemiktarak2012Mechanically, Stambaugh2014From}, readily achieve a linear coupling that is very near the maximum possible for a 100\%-reflective disc, $G_\text{max} = 4\pi c/\lambda L$. In this case, for scattering rate $\Gamma$, the best possible per-photon CM trap efficiency would be 
\begin{equation}\label{eq:ultimate}
K_{\text{ultimate},\Gamma}=16\pi P/\lambda L\Gamma. 
\end{equation}
For these larger perturbations, however, additional cavity modes must be included in the theory (see Appendix III) and it becomes difficult to derive simple analytical expressions describing the avoided crossings. However, the symmetry arguments of Section \ref{sec:general-levitation} remain valid and can be generalized to more modes, meaning the CM and TM gaps will still be tunable (through zero) via $\theta_z$. As discussed in Appendix III, the CM and TM traps should also still have comparable strength. Equation \ref{eq:ultimate} can therefore be used to roughly estimate the expected trap strengths for a variety of geometries. For example, a 110-nm-thick Si disc of radius $r=5$ \um~in a cavity of length $L=100$ \um~\cite{Flowers2012Fiber} and modest finesse $F=1500$, should achieve (for a finesse-limited gap $\Gamma = 2\pi\times 500$ MHz) levitated frequencies $\sim 10$ MHz with only $30$ mW of \emph{circulating} power at $\lambda=1550$ nm. This represents a significantly stronger trap requiring \emph{much} less power than for a single-mode or retro-reflected approach.

Finally, the finesse-limited upper bound discussed in Section \ref{sec:general-quadratic} (Eq.\ \ref{eq:K-Q-limit}) for this geometry can be written 
\begin{equation}
K_{\text{ultimate},F}=32PF/\lambda c.
\end{equation}
Stated briefly, the tilt-control afforded by a second tether provides a means to fully utilize the cavity finesse, without the requirement of engineering a highly-reflective disc.

\section{Torsional Levitation}\label{sec:comsol}
\begin{figure*}[htb]
  \includegraphics[width=1\textwidth]{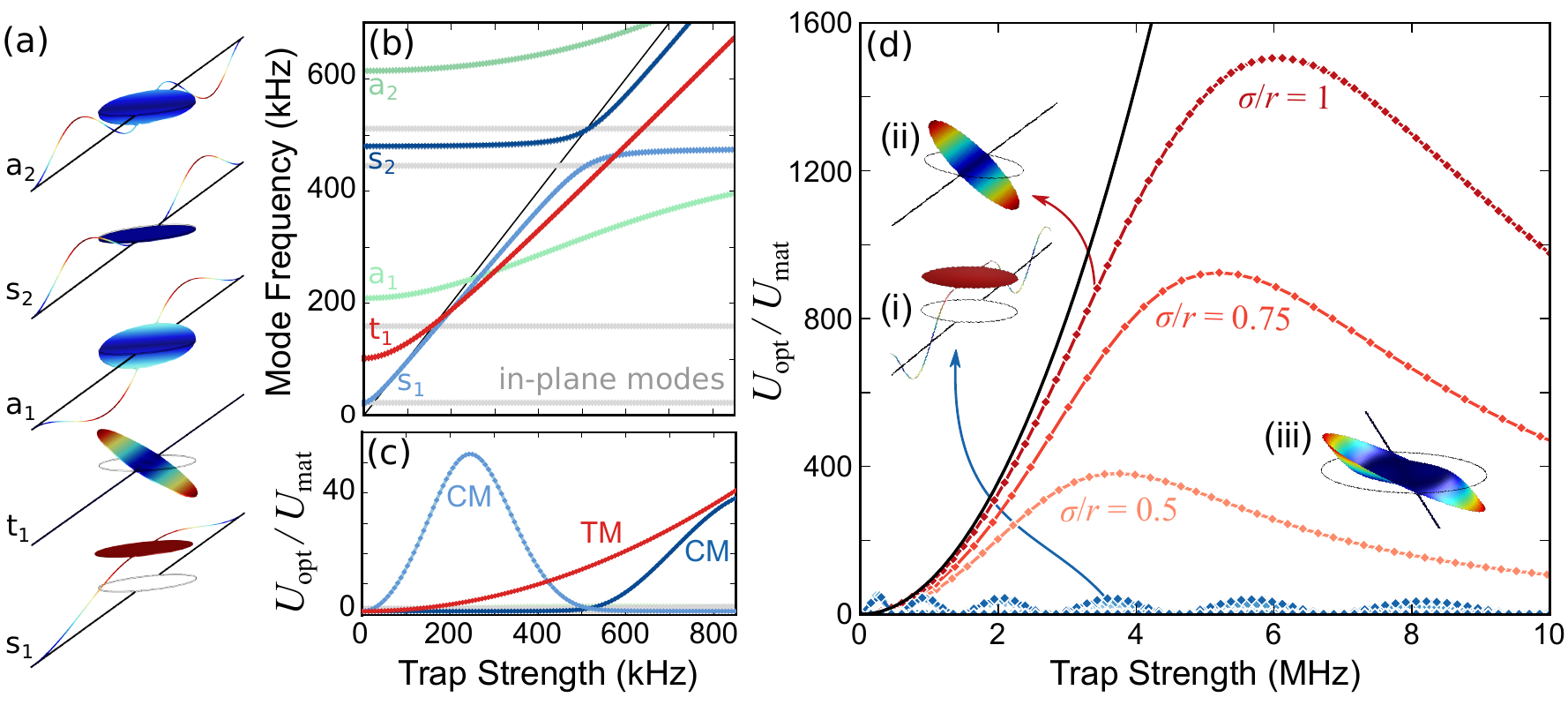}
  \caption{Response of a doubly-tethered silicon disc to a strong optical trap. Structure has thickness $t=110$ nm, disc radius $r = 5$ \um, tether length $l = 45$ \um, and tether width $d= 100$ nm. (a) Relevant low-frequency (untrapped) mechanical modes that are symmetric (``s'') and antisymmetric (``a'') about the $y=0$ plane, or torsional (``t'') (b) Eigenfrequencies versus optical trap strength $\propto \sqrt{P}$. The units of trap strength are scaled to remove the dependence on trapping mechanism (see text). Gray curves correspond to non-interacting in-plane modes. (c) Enhancement $U\opt/U\mat$ for the modes in (b). For CM motion, $U\opt/U\mat$ plummets from hybridization with ``violin'' modes such as $s_2$. (d) Enhancement for the CM-like (blue) and torsional (red) modes at higher trap strength. For each group, the lightest, intermediate, and darkest shades correspond to $\sigma/r = 0.5$, $0.75$ and $1$. The CM mode $Q_m$-enhancement is continuously limited by hybridization with tether modes, (e.g.\ inset i) whereas the torsional hybridization is suppressed (inset ii) until the finite spot size of the trap causes coupling with the ``flappy'' mode of the disc (inset iii). Shown for comparison is the torsional $Q_m$-enhancement for an infinitely stiff structure (black curve). The values of $\sigma/r$ shown here are compatible with high finesse cavities: even $\sigma/r = 1$ could achieve $F>10^4$ for a properly manufactured structure \cite{Chang2012Ultrahigh}.}
  \label{fig:comsol}
\end{figure*}

To investigate how a realistic mechanical element might react to these strong traps, we now discuss a finite-element simulation (COMSOL) of a doubly-tethered disc such as the one in Fig.\ \ref{fig:intro}. The disc and tethers are patterned from a single-crystal silicon sheet of thickness $t=110$ nm, with disc radius $r=5$ \um, tether length $l=45$ \um, and tether width $d=100$ nm. This mechanical element is chosen for its relative ease of fabrication (standard lithography with a silicon-on-insulator wafer), low optical absorption at telecom wavelengths $\lambda=1550$ nm, high reflectivity, low internal stress, and high power handling via a thermal conductivity two orders of magnitude above that of SiO$_2$. The optical trap is modeled as a restoring force along $\hat{x}$ with a spring constant density $K_\text{opt}(y,z) \propto e^{-(y^2+z^2)/2\sigma^2}$, where $\sigma$ is the width of the Gaussian beam \cite{Chang2012Ultrahigh}.

Figure \ref{fig:comsol}(a) shows the relevant (untrapped) mechanical modes. The center of mass ($s_1$), torsional ($t_1$) and antisymmetric ($a_1, a_2$) modes all involve displacement along $\hat{x}$ and are optically trapped, as shown in Fig.\ \ref{fig:comsol}(b). Here, $\sigma = 5$ \um,  and the trap strength is normalized by the response of a perfectly-rigid, tether-free disc in free space (solid black line in Fig.\ \ref{fig:comsol}(b)) to remove the dependence of the result on trap efficiency. Consistent with the aforementioned geometrical considerations, the torsional mode is only slightly less responsive than the CM mode (from the slopes of the linear regions at higher trap strength, $K_{\text{TM}}$ is estimated to be within 10\% of $K_\text{CM}$).

To describe the effect of the optical trap on $Q_m$, we follow Ref.\ \cite{Chang2012Ultrahigh} and parameterize the $Q_m$ enhancement in terms of the potential energy stored in the light field $U\opt \propto K\opt$ and material $U\mat \propto K\mat$. Equation \ref{eq:basic} then predicts $Q_m = Q\mat (K\opt+K\mat)/K\mat \approx Q\mat (U\opt/U\mat)$ for $U\opt \gg U\mat$. This enhancement is shown in Fig.\ \ref{fig:comsol}(c). In agreement with Ref.\ \cite{Chang2012Ultrahigh}, we find that the achievable enhancement depends strongly on the symmetry of the involved mechanical modes. The optical trap hybridizes nominally orthogonal modes of the same symmetry when they are brought into degeneracy, appearing as avoided crossings in the mechanical frequencies and a quenching of $U\opt/U\mat$. For example, the enhancement of the CM (``symmetric'') mode $s_1$ plummets near 500 kHz due to hybridization with the symmetric tether mode $s_2$. A similar avoided crossing between the antisymmetric modes $a_1$ and $a_2$ can also be seen near 600 kHz, where the concavity of the frequency tuning for $a_1$ inverts. Tether mode hybridization represents a fundamental limitation of CM trapping, placing the ceiling $U\opt/U\mat \lesssim 50$ as shown in Fig.\ \ref{fig:comsol}(c) (blue curves). 

Most significantly,the torsional mode does \emph{not} couple with any of these low-frequency modes by symmetry, and $U\opt/U\mat$ increases monotonically in Fig.\ \ref{fig:comsol}(c) (red curve). While the symmetry of a practical structure is never perfect, any residual coupling to the violin modes can in principle be nulled out by aligning the trapping potential to result in a pure twisting of the tethers. Due to its higher initial $\omega\mat$, the overall enhancement is not as immediately large. However, with access to a MHz-scale optical trap, its enhancement can continue far beyond what is possible with the CM mode (or any of the other modes), as shown in Fig.\ \ref{fig:comsol}(d): while the CM mode repeatedly hybridizes with tether modes (e.g.\ inset i), the torsional mode remains essentially unchanged to a much higher frequency (ii).

The eventual ceiling on this enhancement does not necessarily come from the torsional tether modes, because (unlike ``violin string'' modes) their frequencies can be made arbitrarily high by decreasing the tether width; in this case the first torsional tether mode occurs around 57 MHz. Instead, the torsional mode initially hybridizes with the first ``flappy'' mode of the disc (inset iii), which is not as efficiently trapped. In other words, the laser spot strongly pins the center of the disc, leaving the edges relatively free to ``flap''. Figure \ref{fig:comsol}(d) shows simulations performed with $\sigma/r$ = 0.5, 0.75, and 1 (red curves). As the trap is made more uniform across the disc, the flappy modes are trapped more efficiently, and the onset of hybridization occurs at higher frequencies. We find this enhancement factor is fairly insensitive to device thickness, diameter, or choice of material, since these changes affect all modes equally.

In the end, the maximum obtainable $Q_m$ increase for a given geometry involves a trade-off between the trap uniformity, the trap strength per photon, and the degradation of finesse associated with diffraction from the disc. In spite of this, the ratio $\sigma/r = 1$ used here is in principle compatible with a cavity finesse of $10^4$ or larger for a properly engineered disc \cite{Chang2012Ultrahigh}. As shown in Fig.\ \ref{fig:comsol}(d), this corresponds to a $Q_m$ enhancement factor of $\sim 1500$. If the untrapped $Q_m \sim 10^5$-$10^6$, this corresponds to an enhanced $Q_m \sim 10^8$-$10^9$ at a frequency $\sim 10$ MHz. Further improvements can be made at the expense of cavity finesse by further increasing the spot size, and as discussed in Section \ref{sec:specific-levitation} the finesse requirements are not that stringent.

\section{Summary and Discussion}\label{sec:summary}

We have described an efficient optical levitation scheme based on quadratic cavity optomechanical coupling, and described how to realize it with a dielectric disc in a Fabry-Perot cavity. This scheme leads to a strongly enhanced trap for both the center-of-mass and torsional motion of the disc, and simulations suggest that these traps allow the torsional mode to achieve a $Q_m$-enhancement factor far exceeding what is possible with any other mode. Using a trap geometry compatible with a finesse $>10,000$, we predict a $Q_m$-increase of more than three orders of magnitude for a silicon device that can be readily fabricated with standard techniques. This scheme therefore presents a practical platform for a variety of applications ranging from frequency-tuned high-resolution force sensing to quantum optomechanics experiments.

In the above analysis, we focused primarily on how to generate an efficient trap, and how this in turn produces a mechanical element that is highly isolated from the thermal environment of the material. We reiterate that some care must be taken in the cavity design, depending on whether the goal is to enhance the effect of the trap's quantum noise (i.e.\ in the ``bad cavity'' limit $\omega_m \ll \kappa$), or to suppress it ($\omega_m \gg \kappa$). It is also worth noting that while \emph{purely} quadratic optomechanical coupling can be realized with an ideal single-port cavity (see Refs.\ \cite{Thompson2008Strong, Miao2009Standard}), the two optical modes discussed in Section \ref{sec:specific-levitation} ensure the presence of a second port, even if one cavity mirror is perfectly reflective. This can lead to a significant linear contribution to the RPSN \cite{Miao2009Standard}. On the other hand, cavity light landing on the disc's flat surface will tend to scatter back into the cavity mode, which can then be collected and manipulated. It remains an interesting question to what extent the effect of a second port could be interferometrically suppressed, or how RPSN might be further controlled by pre-squeezing the trap light.

\section{Acknowledgments}

We thank Aashish Clerk, Lilian Childress, Marc-Antoine Lemonde, Bogdan Piciu, Alex Bourassa, Simon Bernard, Abeer Barasheed, Xinyuan Zhang, Maximilian Ruf, and Andrew Jayich for helpful discussions and technical support. T.M. acknowledges support by a Swiss National Foundation Early Postdoc Mobility Fellowship. The authors also gratefully acknowledge financial support from NSERC, FRQNT, the Alfred P. Sloan Foundation, CFI, INTRIQ, RQMP, CMC Microsystems, and the Centre for the Physics of Materials at McGill.

\section*{APPENDIX I: Leading-Order Effect of Tilt on $\CCC$ and $\SSS$}\label{app:hermite}

Here we derive the leading-order dependence of $\mathscr{C}_{nm}(\Theta)$ and $\mathscr{S}_{nm}(\Theta)$ on $\Theta$ by expanding $\cos(\Theta\chi)$ and $\sin(\Theta\chi)$ in Eq.\ \ref{eq:curlyCS} and repeatedly applying the Hermite polynomial recursion relation
\begin{equation}\label{eq:recursion}
	\chi H_{n}(\chi) = \frac{1}{2}H_{n+1}(\chi)+nH_{n-1}(\chi)
\end{equation}
to calculate each term.

First, using the ``normalized'' polynomials 
\begin{equation}
\h_n(\chi) = \frac{H_n(\chi)}{\sqrt{2^n n! \pi^{1/2}}}
\end{equation}
of Eq.\ \ref{eq:curlyCS}, the recursion relation can be written
\begin{equation}\label{eq:recursion2}
\chi \h_n(\chi) = \sqrt{\frac{n+1}{2}}\h_{n+1}(\chi)+\sqrt{\frac{n}{2}}\h_{n-1}(\chi).
\end{equation}
This can be used to calculate any term in the series expansion of $\mathscr{C}_{nm}(\Theta)$ or $\mathscr{S}_{nm}(\Theta)$. Each of these terms has the form
\begin{equation}\label{eq:term}
\int (\Theta\chi)^i\h_n(\chi)\h_m(\chi) e^{-\chi^2}d\chi
\end{equation}
with $i,n,m\ge0$ (we also assume $m\ge n$ without loss of generality). Applying Eq.\ \ref{eq:recursion2} $i$ times upon the quantity $\chi^i \h_n(\chi)$ then generates $i+1$ new Hermite polynomials of order at most $n+i$, so for $i<m-n$ all resulting integrals vanish by orthogonality. Similarly, for $i=m-n$ all terms vanish except $\Theta^{i}\sqrt{m!/2^(m-n) n!}\int \h_m^2(\chi)e^{-\chi^2}d\chi$. As a result,
\begin{equation}\label{eq:result}
\hspace*{-0.2cm}\begin{aligned}
\int(\Theta\chi)^{i}\h_{n}\h_{m}e^{-\chi^{2}}d\chi = 
  \begin{cases}
  \Theta^{i}\sqrt{\frac{m!}{2^{m-n} n!}} &\text{if } i=m-n \\
  0 &\text{if } i<m-n
  \end{cases}
\end{aligned}
\end{equation}
Finally, inserting the Taylor series expansion of $\sin(\Theta\chi)$ and $\cos(\Theta\chi)$ into Eq.\ \ref{eq:curlyCS} and incorporating the above result,
\begin{equation}
\hspace*{-0.3cm}\begin{aligned}
\mathscr{C}_{nm}=\CCC_{mn} &\approx \sqrt{\frac{m!}{2^{m-n} n!}}\frac{(-\Theta)^{m-n}}{(m-n)!}, &m-n \text{ even} \\
\mathscr{S}_{nm}=\SSS_{mn} &\approx \sqrt{\frac{m!}{2^{m-n} n!}}\frac{(-\Theta)^{m-n}}{(m-n)!}, &m-n \text{ odd}  
\end{aligned}
\end{equation} 
to leading order for $m>n$, and by symmetry $\mathscr{C}_{nm} = 0$ ($\mathscr{S}_{nm} = 0$) for $m-n$ odd (even) as discussed in the text.

\section*{APPENDIX II: Quartic Traps}\label{app:quartic}

\begin{figure}
	\includegraphics[width=0.48\textwidth]{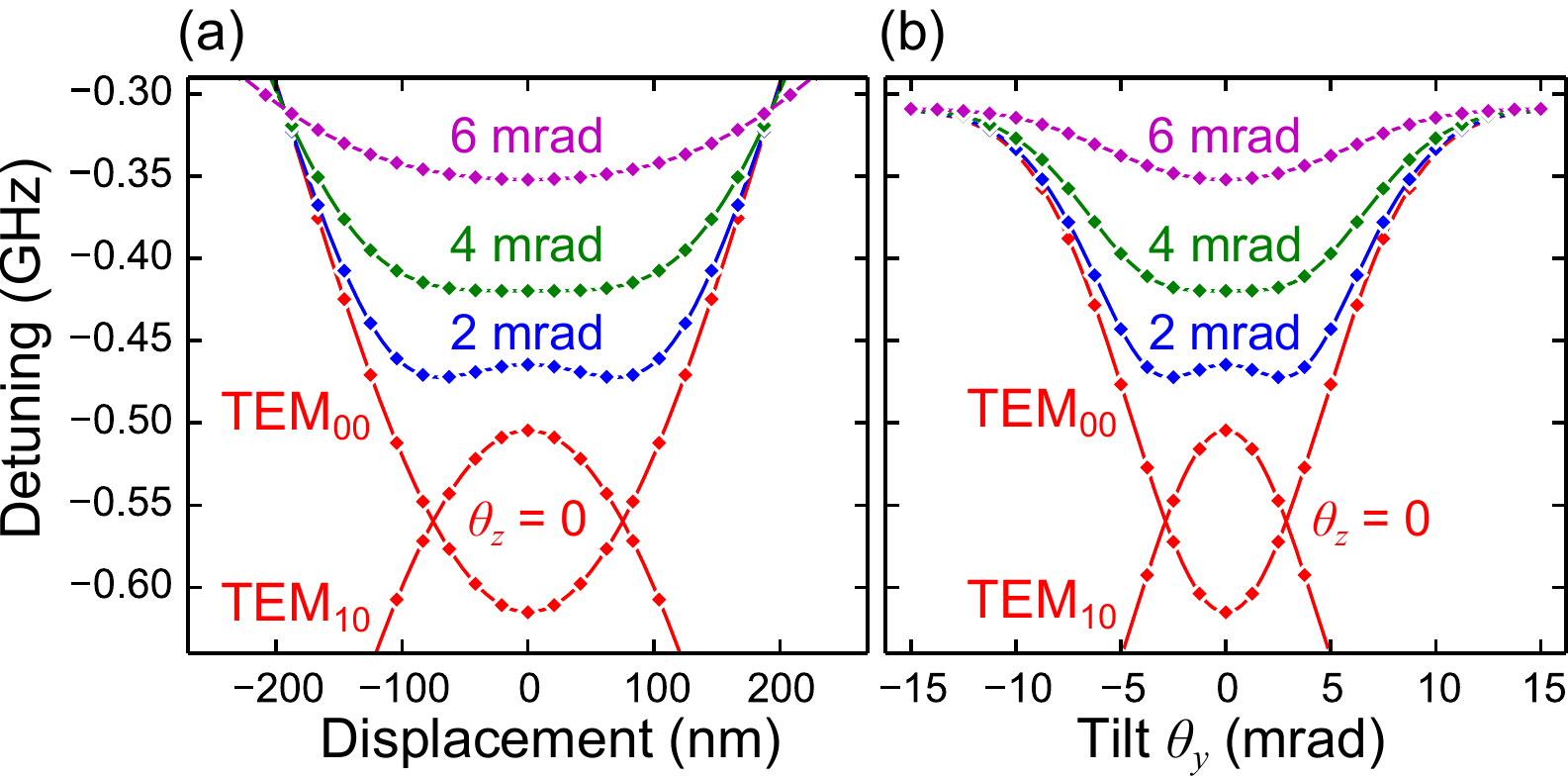}
	\caption{Purely quartic CM and TM coupling. (a) Detuning of the \TEMOO and \TEMIO cavity modes versus displacement of a 50-nm-thick, \SiN disc for a cavity of length $L=4.7$ cm and $\theta_z$ between 0 and 6 milliradians. (b) Detuning versus tilt $\theta_y$ under the same conditions. Both (a) and (b) exhibit a transition from quadratic (6 mrad) to double-well (2 mrad) potentials. In between (near 4 mrad), the optomechanical coupling is purely quartic.}
	\label{fig-quartic}
\end{figure}

As mentioned in Section \ref{sec:specific-levitation}, the expressions for $V_{ij}$ in Eq.\ \ref{eq:Vspecific} can make predictions for a much wider range of tilts than those of previous work \cite{Sankey2009Improved,Sankey2010Strong,Karuza2013Tunable}. As shown in Figure \ref{fig-quartic}, this analytical solution describes how a purely-quartic optomechanical coupling such as the one observed in Ref.\ \cite{Sankey2010Strong} could be generated with two transverse modes of the cavity. Panel (a) shows the detuning of the \TEMOO and \TEMIO modes versus membrane position $x_0$ for more extreme values of $\theta_z$. As $\theta_z$ is increased, the optomechanical coupling near $x_0=0$ for the upper branch makes a smooth transition from double-well ($\theta_z = 2$ milliradians) to quartic ($\theta_z = 4$ milliradians) to quadratic ($\theta_z = 6$ milliradians). The same trend can be observed in (b) for torsional motion.

\section*{APPENDIX III: Stronger Perturbations}\label{app:stronger}

\begin{figure*}[htb]
\includegraphics[width=1\textwidth]{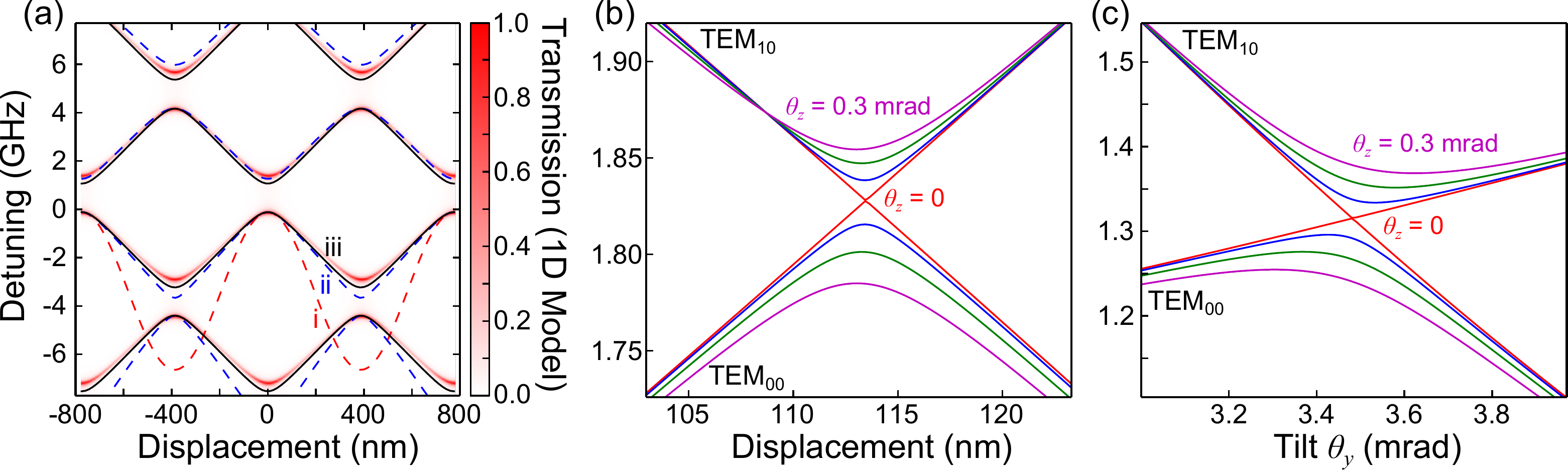}
\caption{Limits of perturbation theory. (a) Detuning of the \TEMOO cavity resonances (near wavelength $\lambda = 1.55 $ \um) for a cavity of length $L=3.5$ cm and mirror radius of curvature 2.5 cm, as a function of disc displacement $x_0$ from the waist. The disc is 110 nm thick, and made of single-crystal silicon (index $n=3.48$), and aligned with the cavity mode ($\theta_y = \theta_z = 0$). Curves show the results of first-order degenerate perturbation theory, including (i) a single \TEMOO mode, (ii) $\pm 2$ \TEMOO modes (i.e.\ 5 in total), and (iii) $\pm 100$ (201 total) \TEMOO modes. The gradient scale shows transmission through the cavity calculated using transfer matrices in 1D for comparison (end mirror amplitude transmission coefficients set to 0.3 for visibility). The qualitative behavior of perturbation theory approaches that of the transfer matrices, however including $\pm 1000$ modes does not noticeably change the result from the solid black curve, hinting that this perturbation is too large to be quantitatively captured by a first-order theory. (b) Displacement-mediated avoided crossings between the \TEMOO and \TEMIO modes, including $\pm 100$ of each type (402 in total), for $\theta_z$ = 0.0, 0.1, 0.2, and 0.3 milliradians. The gap tuning mechanism is roughly linear in $\theta_z$, though a shift in the crossing point is introduced via interactions with adjacent longitudinal modes. (c) Tilt-mediated avoided crossings under the same conditions as (b). The gap tuning is again linear in $\theta_z$, but the crossing modes are no longer symmetrically tuned by $\theta_y$, leading to a skewed crossing. Using this numerically-determined quadratic coupling, a disc of radius $r=75$ \um~(i.e.\ equal to the \TEMOO spot size) experiences a trap frequency ratio $\omega_{\text{TM}}/\omega_{\text{CM}} = 0.43$.}
\label{fig-manymodes}
\end{figure*}

An ideal structure for trapping has low residual stress, high thermal conductivity, and is maximally reflective. As a test material, we selected 110-nm-thick single-crystal silicon, which has an index of $n=3.48$ and is approximately a quarter wavelength thick at $\lambda=1550$ nm. However, such a strongly-perturbing dielectric slab can no longer be treated using a simple first order perturbation theory with a small number of modes.

If we naively solve perturbation theory for such a structure including only one longitudinal \TEMOO mode (i.e.\ the dashed red line in Figure \ref{fig-manymodes}(a)), the problem is immediately apparent: the perturbation is larger than the free spectral range, and inspecting Eq.\ \ref{eq:Vij}, we expect different longitudinal modes with the same transverse profile to strongly scatter into one another. This intuition is confirmed by the complete lack of agreement with a 1D transfer matrix approach (underlying gradient scale). However, this simplified few-mode perturbation theory still quantitatively agrees with the transfer matrix approach so long as the perturbation is sufficiently small compared to the free spectral range. This is why only a few modes are required for the 50-nm-thick silicon nitride disc discussed above.

If we include additional longitudinal modes in the ``nearly degenerate'' manifold, the agreement with the 1D transfer matrix result improves. The blue dashed curve in Fig.\ \ref{fig-manymodes}(a) shows the result of perturbation theory including 5 adjacent longitudinal modes (i.e.\ the mode of interest $\pm 2$ additional modes), and the black curve shows the result for 201 modes ($\pm 100$ modes). At first glance the results seem to be converging to the 1D model, but including, say, 2001 or more modes adds CPU cycles without causing a significant change, implying that this level of perturbation is likely just beyond the grasp of a first-order theory (perhaps not surprisingly). Note that in this calculation we made a point of including the minute differences in mode waist for the longitudinal modes so that the solution could actually converge (albeit incredibly slowly).

Despite the requirement of additional modes, the tilt-based gap-tuning mechanism, which is based on symmetry, should fundamentally remain the same. Figures \ref{fig-manymodes}(b) and (c) show the behavior of avoided gaps in both displacement and tilt including 201 \TEMOO and 201 \TEMIO modes. The additional longitudinal modes introduce some drift in the location of the crossing in (b) and some skew to the linear coupling of the underlying modes in (c), but the overall behavior is otherwise identical: the gap scales roughly linearly with $\theta_z$, and the ratio of trap frequencies $\omega_{\text{TM}}/\omega_{\text{CM}} = 0.43$ for $r=\sigma$ (determined numerically from the shown curves) is of order unity for a given gap.
Scattering between the \TEMOO and higher-order modes of the correct symmetry should be smaller than for the \TEMIO mode, but they should probably still be considered in a more careful analysis.

\bibliography{Alles}

\end{document}